\documentclass[seceq]{ptptex}
\newcommand{\siml}{\hspace{.3em}\raisebox{.4ex}{$<$}\hspace{-.75em}
 \raisebox{-.7ex}{$\sim$}\hspace{.3em}}
\newcommand{\simg}{\hspace{.3em}\raisebox{.4ex}{$>$}\hspace{-.75em}
 \raisebox{-.7ex}{$\sim$}\hspace{.3em}}
\usepackage{graphicx}

\preprintnumber[4.5cm]{
SAGA-HE-220\\YAMAGATA-HEP***\\ June, 2005}

\title{
True or Fictitious Flattening? \par
 --MEM and  the $\theta$ Term  --
}

\author{
 Masahiro Imachi$^{\ddagger}$\footnote{E-mail: imachi@sci.kj.yamagata-u.ac.jp},\\
  Yasuhiko Shinno$^{\diamond}$\footnote{E-mail: 
  shinno@dirac.phys.saga-u.ac.jp} and 
   Hiroshi Yoneyama$^{\diamond}$\footnote{E-mail: yoneyama@cc.saga-u.ac.jp}
}

\inst{
 Department of Physics, Yamagata University, Yamagata 990-8560, Japan$^{\ddagger}$\\
 Department of Physics, Saga University, Saga 840-8502, Japan$^{\diamond}$\\}

\abst{
We study the sign problem in lattice field theory with a  $\theta$ 
term.   We apply the maximum 
 entropy method (MEM)  to flattening phenomenon  of  the free 
energy density $f(\theta)$, which originates from   the sign problem.  
In our  previous paper,  we applied  the MEM  by employing  the  Gaussian topological charge 
distribution  $P(Q)$ as mock data.
In the present paper,  we consider models in  which   `true'  flattening of $f(\theta)$ occurs. 
 These  may be   regarded as good examples for studying whether the MEM could 
correctly detect non trivial phase structure. 
}

\begin{document}
\maketitle
\section{Introduction}
\label{sec:introduction}
\par
A  $\theta$  term in QCD is expected to cause interesting dynamics.~\cite{rf:tHooft,rf:CR}  \ 
Lattice field theory  may be a suitable tool as a non perturbative method to study its
 effect.~\cite{rf:Seiberg,rf:Wiese,rf:HITY,rf:BISY,rf:FO}   \  However, 
 it suffers from the complex action
    problem or  the sign problem,  when  a  $\theta$  term is included.
       A conventional technique to avoid the problem is to perform the Fourier
   transform of the topological charge distribution
  $P(Q)$, which is calculated with real positive Boltzmann weight,  and
  to calculate the partition function ${\cal Z}(\theta)$.~\cite{rf:BRSW, rf:Wiese2}  \  However, this still
  causes  flattening of the free energy density $f(\theta)$ originated from the error in $P(Q)$.~\cite{rf:PS,rf:IKY} \ 
  This  phenomenon  misleads  into a  fictitious signal of a first order phase transition.~\cite{rf:OS}  \ 
  Some  approaches have been  tried to overcome the sign problem.~\cite{rf:ACGL,rf:AANV}  \  
     In our previous paper~\cite{rf:ISY} (referred to as (I) hereafter), we applied  to this issue
      the maximum entropy method (MEM), which 
   is suitable to deal with so-called ill-posed problems, where the number of parameters to be 
     determined is much larger than the number of data points. 
           The MEM is based upon  Bayes' theorem. It derives the most probable
  parameters by utilizing data sets and our knowledge about these
  parameters in terms of the probability.~\cite{rf:Bryan,rf:SSG,rf:GJSS,rf:JGU,rf:AHN}  \ The probability distribution,
  which is called posterior probability, is given by the product of the
  likelihood function and the prior probability. The latter is
  represented by the Shannon-Jaynes entropy, which plays an important
  role to guarantee the uniqueness of the solution, and the former is
  given by $\chi^2$. The task is to explore the image such that the
  posterior probability is maximized.      
 \par
  In the paper (I),  we used the Gaussian $P(Q)$ and  tested
   whether the  MEM would be effective, since the Gaussian $P(Q)$ is
  analytically Fourier-transformed to ${\cal Z}(\theta)$.
   For the analysis, we used mock data by  adding Gaussian noise to $P(Q)$. The
  results we obtained are the followings;
  In the case without  flattening, the results of the MEM agree with 
        those  of the Fourier transform method and thus reproduce the exact
        results. In the case with  flattening, the MEM gives
  smoother  $f(\theta)$  than that of the  Fourier transform.  Among
   various default models investigated, some images with the  least
  errors do not show  flattening. \par
   If the reason for the disappearance of  flattening in the above analysis is the smoothing
  of the data, which is  the characteristic of the parameter inference
  such as  the maximal likelihood method, one may wonder if  the MEM may not be 
  a suitable method for detecting  singular behaviors such as a 
  phase transition. 
 In order to investigate   this issue, we study whether  the MEM would be applicable to 
  `true'  flattening of $f(\theta)$ or 
 singular behaviors  like    a  first order phase  transition       occurring at a finite value of $\theta (\neq \pi)$.  By   `true',  it is meant that flattening is  not originated  from the error of  $P(Q)$
 but occurs due to the data themselves.
  For this purpose, we consider a simple  model based on  mock data.  
It  is  a model which consists of the
 Gaussian    $P(Q)$ except at  $Q=0$.  For  $Q=0$, a constant is added to the Gaussian $P(Q)$.
 This  additional contribution at $Q=0$  causes the flattening behavior in the  large $\theta$ region ($\theta \leq \pi$).  
  We apply the MEM to this model and study  whether true 
  flattening would be properly reproduced. \par
  Although this model develops true flattening,  it     does not reveal the singularity in   $f(\theta)$ 
  at finite value of $\theta (\neq \pi)$.
  In order to study  such a  singular behavior mimicking  a  first order phase transition,  we consider  
  a model which utilizes $P(Q)$ obtained by  the inverse Fourier transform of  a  singular $f(\theta)$. 
 In order for the singularity  and flattening to  be  compatible in such a model, obtained    $P(Q)$  oscillates  and could take negative values for large values of $Q$.
 Although  $P(Q)$ looses the physical meaning as the topological charge distribution for their 
  negative values,  we regard this as  a mathematical model  to study the singular behavior in  terms of the MEM. \par 
  We apply the MEM to these  models and  study  (i) whether true 
  flattening would be properly reproduced and  (ii)   how  the errors in $P(Q)$ are
   related  with the results of the MEM and the 
numerical Fourier transform method.
  Although these models  are simple,  we expect that this analysis  would  serve to obtain knowledge if the MEM 
  could be utilized  to study the phase structure,  when it were  applied to other 
  models suffering from the sign problem, say,   finite density QCD. 
\par
   This paper is organized as follows. In the following section, we
   give an   overview of the origin of  flattening and a sketch of the MEM.  
   A  model developing true flattening is also introduced.  
   In $\S$~\ref{sec:toymodel}, the MEM
   is applied to this  model.  In $\S$~\ref {sec:singularf},  another model developing  a singular behavior  is introduced and analyzed.
   \ A summary is given  in
  $\S$~\ref{sec:summary}. \par

\section{MEM and a model}
\label{sec:flattening&MEM}\par
\subsection{Flattening of $f(\theta)$}
\label{sec:flattening}\par
Let us briefly explain flattening of $f(\theta)$.  
   The partition function
${\cal Z}(\theta)$ can
  be obtained by Fourier-transforming the topological charge distribution
  $P(Q)$, which is calculated with real positive Boltzmann weight.
\begin{equation}
  {\cal Z}(\theta)=\frac{\int[d\bar{z}dz]e^{-S+i\theta\hat{Q}(\bar{z},z)}}
   {\int[d\bar{z}dz]e^{-S}}\equiv\sum_{Q}e^{i\theta Q}P(Q),
    \label{eqn:partitionfunction}
\end{equation}
  where
\begin{equation}
  P(Q)\equiv \frac{\int[d\bar{z}dz]_Qe^{-S}}{\int d\bar{z}dz e^{-S}}.
   \label{eqn:Pq}
\end{equation}
  The measure $[d\bar{z}dz]_Q$ in Eq.~(\ref{eqn:Pq}) represents that the
integral is
  restricted to configurations of the field $z$ with the topological
charge $Q$, and
  $S$ denotes an action. \par
  The distribution $P(Q)$ obtained from  Monte Carlo  simulations can be
   decomposed into two parts, a true value ${\tilde P}(Q)$ and an error of $P(Q)$.
When the  error at $Q=0$ dominates, the free energy density becomes 
\begin{eqnarray}
   f(\theta)&=&-\frac{1}{V}\log{\cal Z}(\theta) \approx
    -\frac{1}{V}\log\bigl\{\sum_{Q}{\tilde P}(Q)e^{i\theta Q}+\delta
P(0)\bigr\}\nonumber\\
   &=&-\frac{1}{V}\log\bigl\{e^{-V{\tilde f}(\theta)}+\delta P(0)\bigr\},
    \label{eqn:freeenergydata}
\end{eqnarray}
where ${\tilde f}(\theta)$ is the true free energy density and $\delta P(0)$ is the error at $Q=0$.
   The value of $e^{-V{\tilde f}(\theta)}$ decreases rapidly as the volume $V$
   and/or $\theta$ increases. When  the magnitude of
   $e^{-V{\tilde f}(\theta)}$ becomes comparable to that of  $\delta P(0)$ at $\theta\simeq\theta_{\rm f}$, 
   $f(\theta)$ becomes almost constant,   $f(\theta)\simeq{\rm const}$,  for
   $\theta\simg\theta_{\rm f}$. This abrupt change of the behavior is  misleadingly identified as   
   a first order phase transition at  $\theta\approx \theta_{\rm f}$.
   This  is (fictitious)  flattening phenomenon  of $f(\theta)$.             
 The error in $P(Q)$ could  also lead to negative             values  of ${\cal Z}(\theta)$. 
We   refer  to  this, too,     as flattening,  because its  origin is   the same.  
In the case of a large volume, an exponentially increasing amount of data is needed.\par
\subsection{MEM}
\label{sec:MEM}
In this subsection, we  briefly explain the MEM in the context of the $\theta$ term. 
 See (I) for details.
    Instead of dealing with Eq.~(\ref{eqn:partitionfunction}), we consider
\begin{equation}
   P(Q)=\int_{-\pi}^{\pi}d\theta \frac{e^{-i\theta Q}}{2\pi}{\cal Z}(\theta).
    \label{eqn:Pqint}
\end{equation}
The MEM is based on  Bayes' theorem. In order
to calculate ${\cal Z}(\theta)$,  we maximize the probability
\begin{equation}
   {\rm prob}({\cal Z}(\theta)|P(Q),I)={\rm prob}({\cal Z}(\theta)|I)
    \frac{{\rm prob}(P(Q)|{\cal Z}(\theta),I)}{{\rm prob}(P(Q)|I)},
    \label{eqn:BayestheoremforPq}
\end{equation}
where ${\rm prob}(A)$ is the probability that an
   event $A$ occurs, and ${\rm prob}(A|B)$ is the conditional probability
   that $A$ occurs under the  condition that  $B$ occurs.

      The likelihood function ${\rm prob}(P(Q)|{\cal Z}(\theta),I)$ is given by
\begin{equation}
   {\rm prob}(P(Q)|{\cal Z}(\theta),I)=\frac{e^{-\frac{1}{2}\chi^{2}}}{X_L},
    \label{eqn:likelihoodfunction}
\end{equation}
   where $X_L$ is
   a normalization constant, and $\chi^{2}$  is represented
   by
\begin{equation}
   \chi^2\equiv \sum_{Q,Q^\prime}(P^{({\cal Z})}(Q)-{\bar P}(Q))
    C^{-1}_{Q,Q^\prime}
    (P^{({\cal Z})}(Q^\prime)-{\bar P}(Q^\prime)).
    \label{eqn:chisquare}
\end{equation}
 Here,  $P^{({\cal Z})}(Q)$ is constructed from
   ${\cal Z}(\theta)$ through Eq.~(\ref{eqn:Pqint}), and  ${\bar
   P}(Q)$ denotes the average of a data set $\{P(Q)\}$;
\begin{equation}
   {\bar P}(Q)=\frac{1}{N_d}\sum_{l=1}^{N_d}P^{(l)}(Q), \label{eqn:average}
\end{equation}
   where $N_d$ represents the number of sets of  data. The matrix
   $C^{-1}$ represents the inverse covariance matrix obtained  by the  data
   set    $\{P(Q)\}$.
   \par
   The prior probability ${\rm prob}({\cal Z}(\theta)|I)$ is represented in
   terms of the entropy $S$ as
\begin{equation}
   {\rm prob}({\cal Z}(\theta)|I,\alpha)=\frac{e^{\alpha S}}{X_S(\alpha)},
    \label{eqn:priorprobability}
\end{equation}
   where $\alpha$ is a real positive parameter and $X_S(\alpha)$ denotes
   an $\alpha$-dependent normalization constant. With regard to  $S$,
   the Shannon-Jaynes entropy is employed:
\begin{equation}
   S=\int^\pi_{-\pi} d\theta \biggl[{\cal Z}(\theta)-m(\theta)-
    {\cal Z}(\theta)\log\frac{{\cal Z}(\theta)}{m(\theta)}\biggr],
    \label{eqn:SJentropy}
\end{equation}
   where $m(\theta)$ is called the `default model'.
      The  posterior probability  thus amounts to
\begin{equation}
   {\rm prob}({\cal Z}(\theta)|P(Q),I,\alpha,m)=\frac{e^{-\frac{1}{2}
    \chi^{2}+\alpha S}}{X_L X_S(\alpha)},
    \label{eqn:posteriorprobability}
\end{equation}
   where it is explicitly shown that $\alpha$ and $m$ are regarded as new
   prior knowledge in ${\rm prob}({\cal Z}(\theta)|P(Q),I,\alpha,m)$.
   \par
   For the  information $I$,  we impose the  criterion
\begin{equation}
   {\cal Z}(\theta)>0 \label{eqn:criterion}
\end{equation}
   so that ${\rm prob}({\cal Z}(\theta)\leq 0|I,m)=0$.
   \par
     We employ  the following  procedure for the  analysis.~\cite{rf:Bryan,rf:AHN} \ 
   Details are given in (I).
\vspace*{4mm}
\begin{enumerate}
   \item Maximizing $W$  to obtain ${\cal Z}^{(\alpha)}(\theta)$ for a fixed $\alpha$: \par
\hspace*{5mm}
         In order to find the image in functional
         space of ${\cal Z}(\theta)$ for a given $\alpha$, we maximize 
\begin{equation}
   W\equiv -\frac{1}{2}\chi^{2}+\alpha S
    \label{eqn:functionw}
\end{equation}         
so that 
\begin{equation}
   \frac{\delta}{\delta{\cal Z}(\theta)}(-\frac{1}{2}\chi^2+\alpha S)
    \Bigm|_{{\cal Z}={{\cal Z}^{(\alpha)}}}=0. \label{eqn:maximumcondition}
\end{equation}         
   The parameter $\alpha$ plays the  role determining the relative weights
   of   $S$.  \par
   In the numerical analysis, the continuous function ${\cal Z}(\theta)$
   is discretized;\\
   ${\cal Z}(\theta)\to{\cal Z}(\theta_n)\equiv{\cal Z}_n$. The integral
   over $\theta$ in Eq.~(\ref{eqn:Pqint}) is converted to the finite
   summation over $\theta$;
\begin{eqnarray}
   P_j=\left\{
       \begin{array}{ll}
        \displaystyle{\sum_{n=1}^{N_\theta}\frac{1}{2\pi}{\cal Z}_n}
	\;\;\;\;\;\;\;(j=1) \\
        \displaystyle{\sum_{n=1}^{N_\theta}\frac{\cos\theta_n j}{\pi}
	{\cal Z}_n}\;(\mbox{otherwise})
       \end{array}
      \right\}
   \equiv\sum_{n=1}^{N_\theta}K_{jn}{\cal Z}_n,  \label{eqn:Pqintdis}
\end{eqnarray}
   where $j=1,2,\cdots, N_q$ and $n=1,2,\cdots,  N_{\theta}$.  Note that    $P_j$ denotes  $P(Q)$ at $Q=j-1$ and $N_q< N_\theta$.    Here we used the fact that $P(Q)$ and ${\cal
  Z}(\theta)$ are even functions of $Q$ and $\theta$, respectively.   \par
 \vspace*{4mm}
\item  Averaging ${\cal Z}^{(\alpha)}_n$: \par
\hspace*{5mm}
         The $\alpha$-independent final image, denoted as $ {\hat {\cal Z}}_n$ or ${\hat {\cal Z}}(\theta)$, 
         can be calculated by averaging the image ${\cal Z}^{(\alpha)}_n$		 
	according to the probability. 
\begin{equation}
   {\hat {\cal Z}}_n=\int d\alpha~{\rm prob}(\alpha|P(Q),I,m){\cal Z}^{(\alpha)}_n.
    \label{eqn:averageofZ2Pa}
\end{equation}
   The probability  ${\rm prob}(\alpha|P(Q),I,m)$ is given by 
	\begin{equation}
  {\rm prob}(\alpha|P(Q),I,m)=\frac{1}{X_W}\exp\biggl\{\Lambda(\alpha)+W({\cal
     Z}^{(\alpha)})\biggr\},
     \label{eqn:averageofZ2}
\end{equation}
   where $X_{{\rm W}}$ is a normalization constant, and
   $\Lambda(\alpha)\equiv
         \frac{1}{2}\sum_k\log\frac{\alpha}{\alpha+\lambda_k}$.
  	Here   the values $\lambda_k$'s are eigenvalues of the real symmetric
         matrix in $\theta$ space;
         \begin{equation}
   \frac{1}{2}\sqrt{{\cal Z}_m}
    \frac{\partial^2\chi^2}{\partial{\cal Z}_m\partial{\cal Z}_n}
     \sqrt{{\cal Z}_n}\Bigm|_{{\cal Z}={\cal Z}^{(\alpha)}}.
     \label{eqn:matrixlambda}
\end{equation}
	\par
         In averaging over $\alpha$, we determine a range of $\alpha$ so
         that
         ${\rm prob}(\alpha|P(Q),I,m)\geq  \frac{1}{10} \times  {\rm prob}({\hat
\alpha}|P(Q),I,m)$
         holds, where ${\rm prob}(\alpha|P(Q),I,m)$ is maximized at
         $\alpha={\hat \alpha}$.
\vspace*{4mm}

\item Error estimation~\cite{rf:JGU,rf:AHN}:\par
       The error  of the final output image ${\hat {\cal Z}}_n$
         is calculated based on the uncertainty  of the image, which takes into account the correlations of  the images ${\hat {\cal Z}}_n$ among  different values of $n$.~\cite{rf:ISY}
\end{enumerate}
\vspace*{4mm}
\par
\subsection{A model}
\label{sec:subtoymodel}\par
In (I), we used the Gaussian $P(Q)$ for the MEM analysis. 
    We parameterized the Gaussian $P(Q)$ as 
\begin{equation}
   P_{\rm G}(Q)\equiv A \exp[-\frac{c}{V}Q^2] ,
    \label{eqn:Pqparametrize}
\end{equation}
  where, e.g., in the case of the 2-d U(1) gauge model, $c$ is a constant
   depending on the inverse coupling constant $\beta$,  and $V$ is the lattice
   volume. Here $V$ is regarded as a parameter and varied in the
analysis.
   The  constant  $A$ is  fixed  so that   $\sum_Q P_{\rm G}(Q)=1$.
In order to study the question  whether  the MEM could  be a useful tool 
for reproducing  true flattening, 
we consider a simple  model.\par
Suppose in some lattice theory that calculated $P(Q)$ 
were  slightly deviated from the Gaussian one only at $Q=0$:
\begin{equation}
  P(Q)=P_{\rm G}(Q)+p_0 \times \delta_{Q 0}, 
\label{eqn:Pqa}
\end{equation}
where $p_0$ is a positive constant  and $ \delta_{Q 0}$ is the  Kronecker   symbol. 
    The partition function is analytically obtained  by use of the
Poisson sum formula as
\begin{equation}
   {\cal Z}_{\rm pois}(\theta)=A'\biggl\{\sqrt{\frac{\pi V}{c}}\sum_{n=-\infty}^{\infty}
    \exp\biggl[-\frac{V}{4c}(\theta-2\pi n)^2\biggr] +p_0 \biggr\},\label{eqn:poissonsum}
\end{equation}
where $A'$ is the normalization constant, which is   determined so that $ {\cal Z}_{\rm pois}(0)=1$.
Because the first term in Eq.~(\ref{eqn:poissonsum}) is monotonically
decreasing function of $\theta$,
the role  of the constant term is to cause  a flat distribution
 at large values of $\theta$  ($\theta\leq\pi$). Let us call this model a flattening model.
  We use the distributions $P(Q)$, Eq.~(\ref{eqn:Pqa}),    to generate mock data by adding 
   Gaussian noises to them  as described below. \par
\subsection{Mock data}
\label{sec:mock}\par
       For preparing the mock data, we add noise with the variance of
   $\delta \times P(Q)$ to   $P(Q)$.
 This is based on the
observations  in the procedure which was employed  to calculate
$P(Q)$ in the simulations of  models such as the CP$^{N-1}$
model.~\cite{rf:HITY} \ 
This turns out to yield  a ratio of the error  to the data to be almost constant;
$|\delta P(Q)/P(Q)|\approx$ const.
The  parameter $\delta$ directly affects  the covariance
matrix, and indirectly influences  the error $\delta \hat {{\cal Z}}(\theta)$
of the image, which
is  the uncertainty of the image.  As done in (I),
$\delta$ is mainly fixed to  1/400.   In order to investigate the effects of $\delta$ on  the final  image,  we also use its  smaller values. \par
 A set of data consists of $P(Q)$
with the errors for $Q=0$ to $N_{q}-1$.
Employing   $N_d$ such sets of
data, we calculate the covariance matrices in
Eq.(\ref{eqn:chisquare}) with  the jackknife method as
\begin{equation}
   C_{Q,Q^\prime}=\frac{1}{N_d ( N_d-1)}\sum_{l=1} ^{N_d}(
P^{(l)}(Q)-{\bar P}(Q))( P^{(l)}(Q^\prime)-{\bar P}(Q^\prime) ),
   \label{eqn:covariance}
\end{equation}
  where $P^{(l)}(Q)$ denotes the $l$-th data of the topological charge
distribution and ${\bar P}(Q)$ is the average Eq.(\ref{eqn:average}).
The inverse covariance matrix is calculated with precision
   that the  product of  the   covariance matrix  and its inverse has
off-diagonal elements at most ${\cal O}(10^{-23})$.
 We take    $N_d=30$ as in (I). \par
The number of degrees of freedom in $\theta$ space,  $N_\theta$, is
taken to be larger than that of the topological charge $N_q$:
$N_\theta$ is set to be 28.~\cite{rf:ISY} \ 
The number $N_q$  is chosen so as to satisfy  $P(Q)\geq 10^{-20}$ and to satisfy the quality of 
the covariance matrix described above.  It   varies  depending on which type of model is used and  on $V$.
Note that in order to reproduce ${\cal Z}(\theta)$ which ranges  many
orders,  the analysis  must be  performed with  quadruple  precision.
In order to find a unique solution in the huge configuration space of the image, we employ 
the singular value decomposition.
\par
  As to the default model $m(\theta)$ in Eq.(\ref{eqn:SJentropy}), we
study   various  cases:
       (i) $m(\theta)=$ const. and (ii) $m(\theta)=\exp(-\frac{\ln 10}{\pi^2}  \gamma
\theta^2) $.  In the case of (i), we show the results of 
$m(\theta)= 1.0$.
 The case (ii) is the
       Gaussian default, where the parameter $\gamma$ is  varied in the
analysis. \par
     In the next section, we investigate  the  following questions:
 (i) Does the MEM reproduce true flattening?    
(ii)   How does  the final image of
    the partition function depend on $p_0$  as well
    as   on  the magnitude of the error in $P(Q)$?

\section{Results}
\label{sec:toymodel}\par
%
\subsection{Flattening model}
\begin{figure}
        \centerline{\includegraphics[width=10 cm, height=8
cm]{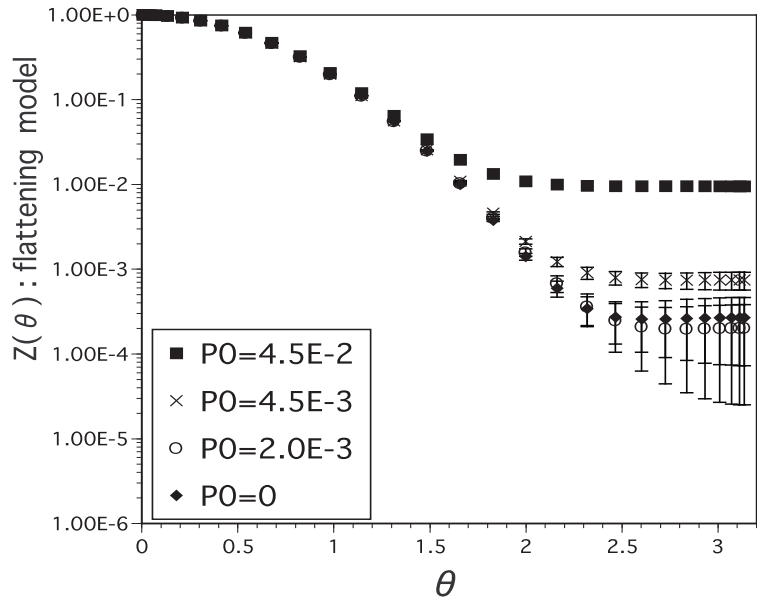}}
\caption{Partition function  ${\cal Z}(\theta)$ calculated by numerical Fourier transform
for $V=50$. Data for $P(Q)$ in Eq.~(\ref{eqn:Pqa}) are used
with $p_0=4.5\times 10^{-2}$, $4.5\times 10^{-3}$, $2.0\times 10^{-3}$ and $0$.  
Here,  $\delta=1/400$.
 Clear flattening is observed in each 
case.  The result  for   $p_0=0$ looks similar to that for $2.0\times 10^{-3}$. The latter 
indicates  somewhat smaller  errors than the former.}
\label{fig:ZV50c0&}
\end{figure}
%
\begin{figure}
        \centerline{\includegraphics[width=10 cm, height=8
cm]{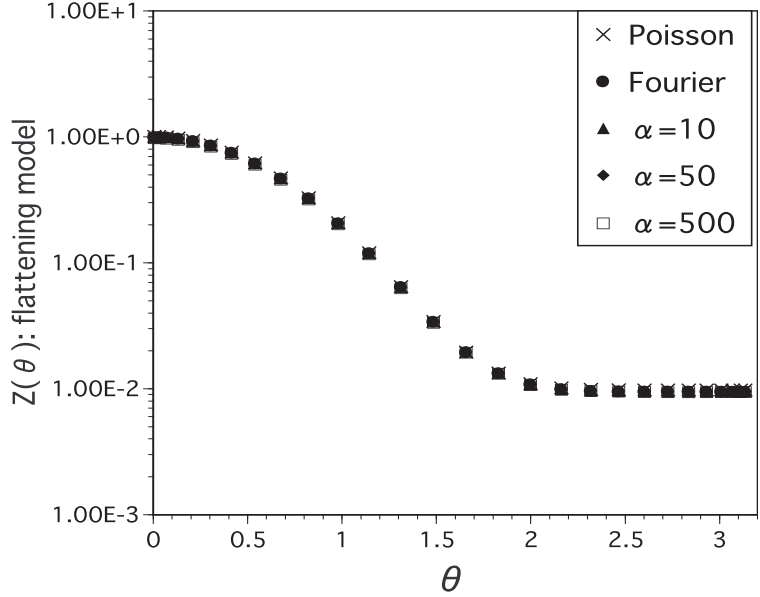}}
\caption{${\cal Z}^{(\alpha)}(\theta)$ obtained in the procedure 1, Eq.~(\ref{eqn:maximumcondition}),  for various values of $\alpha$ in the  flattening model.
Here,  $V=50$, $p_0=4.5\times 10^{-2}$ and $\delta=1/400$. The default model is chosen to be the constant 
 $m(\theta)=1.0$. 
 }
\label{fig:ZV50c4e-2}
\end{figure}
 When  noise is added, the partition function for  Eq.~(\ref{eqn:Pqa}) in the flattening model
  is given by
\begin{equation}
   {\cal Z}(\theta)= \frac{\bigl\{\sum_{Q} P_{\rm G}(Q)e^{i\theta Q}+p_0+
   \sum_{Q} \delta P(Q)e^{i\theta Q}\bigr\}}{B},\label{eqn:toyerror}
\end{equation}
where $B=\sum_{Q} \bigl\{ P_{\rm G}(Q) +p_0\delta_{Q0} + \delta P(Q) \bigr\}$.
Figure~\ref{fig:ZV50c0&} displays the behavior of ${\cal Z}(\theta)$  calculated by
 the numerical Fourier transform of   Eq.~(\ref{eqn:Pqa})
 for $V=50$ and  for various values of $p_0$; $p_0=4.5\times 10^{-2}$, $4.5\times 10^{-3}$ and $2.0\times 10^{-3}$. The parameter   $c$ in Eq.~(\ref{eqn:Pqparametrize}) is fixed  to   7.42  throughout  the paper.   The parameter $\delta$ is chosen to be  $1/400$. 
\begin{figure}
        \centerline{\includegraphics[width=9 cm, height=7
cm]{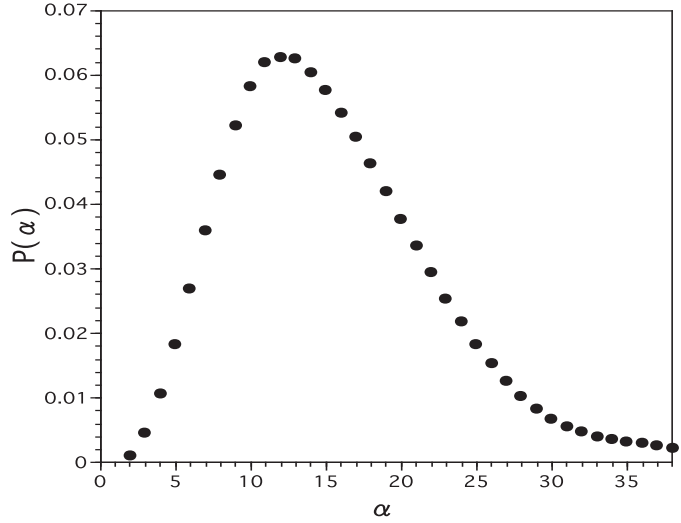}}
\caption{Probability ${\rm prob}(\alpha|P(Q),I,m) \equiv P(\alpha)$
for $V=50$, $p_0=4.5\times 10^{-2}$ and $\delta=1/400$  in the  flattening model. 
The default model is chosen to be the constant 
  $m(\theta)=1.0$.
 }
\label{fig:PV50c4e-2}
\end{figure}
\begin{figure}
        \centerline{\includegraphics[width=9cm, height=7
cm]{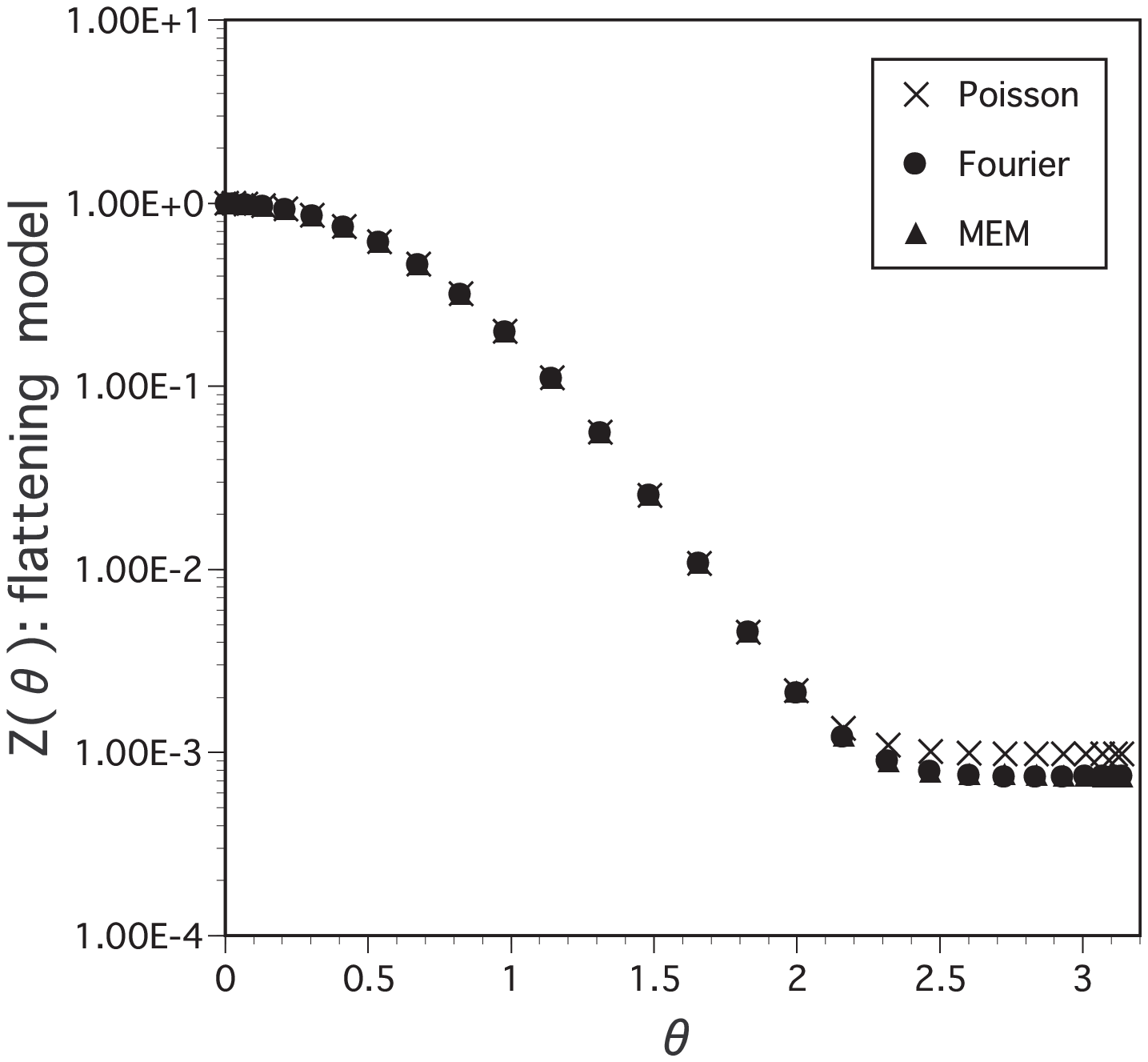}}
\caption{${\cal Z}(\theta)$
for  $V=50$, $p_0=4.5\times 10^{-3}$ and $\delta=1/400$ in the  flattening model. The default model is chosen to be the constant  
 $m(\theta)=1.0$.  Error bars are also  displayed  but invisible.
 }
\label{fig:ZV50c4e-3}
\end{figure}
  Figiure~\ref{fig:ZV50c0&} also displays  ${\cal Z}(\theta)$ for
$p_0=0$, which was studied in (I) in detail.
 Clear  flattening is observed in each  case. 
 Note that the behavior of 
 ${\cal Z}(\theta)$  and its error  are  similar for  the $p_0=2.0\times 10^{-3}$ and $p_0=0$ cases.  
   In order to
understand   what the origin of  these flattening is, let us investigate
Eq.~(\ref{eqn:toyerror}) briefly:
When $p_0$ is so   chosen  that for small values of
$\theta$,
the first term dominates in the
numerator, while for large  values of $\theta$,
the constant term dominates,
 the resultant $ {\cal Z}(\theta)$ behaves like the partition function
 with  flattening. For the latter case,
  Eq.~(\ref{eqn:toyerror})
is approximated by a constant
\begin{equation}
 {\cal Z}(\theta) \approx \frac{p_0}{B}.
  \label{eqn:cB}
\end{equation}
For  $V=50$ and $p_0=4.5\times 10^{-2}$ , $B$ is
 numerically estimated as 4.45 and thus
 \begin{equation}
 {\cal Z}(\theta) \approx \frac{4.5\times
10^{-2}}{4.45}=1.0\times 10^{-2} .
\end{equation}
For other choice of $p_0$, $4.5\times 10^{-3}$ and $2.0\times 10^{-3}$, $p_0/B$ becomes 
$1.0\times 10^{-3}$ and $4.5\times 10^{-4}$, respectively.
These values agree with those  of flattening  observed in  Fig.~\ref{fig:ZV50c0&},
 and we thus find that these  flattening are  not caused by  the error
 in $P(Q)$ but  are due to the additional  term in  Eq.~(\ref{eqn:Pqa}). 
On the other hand, to  flattening for the  $p_0=0$ case  in Fig.~\ref{fig:ZV50c0&},
 Eq.~(\ref{eqn:cB})  does not apply obviously.  What is  observed is not caused by 
the constant term but  
  comes from  the error  in $P(Q)$ as described in section  \ref{sec:flattening}.
 The former  type of flattening  is  true flattening,  because  its  origin  lies in 
 data of $P(Q)$, while the latter ($p_0=0$) is  fictitious one, because it comes from the error in $P(Q)$.  
  \par
%
 \begin{figure}
        \centerline{\includegraphics[width=9cm, height=7
cm]{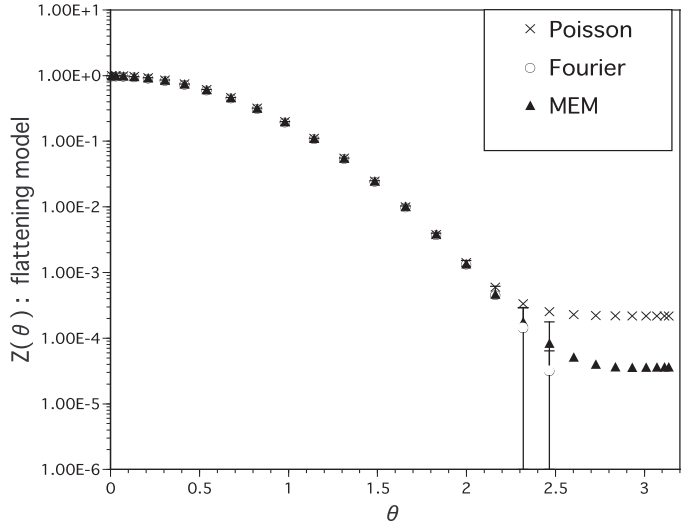}}
\caption{${\cal Z}(\theta)$
for  $V=50$, $p_0=1.0\times 10^{-3}$ and $\delta=1/400$ in the  flattening model. The default model is chosen to be the constant 
 $m(\theta)=1.0$.   Error bars are also  displayed  but invisible for the MEM results.
 The Fourier method yields   negative values for ${\cal Z}(\theta)$   at  
   $\theta \geq 2.5$. 
 }
\label{fig:ZV50c1e-3}
\end{figure}
\begin{table}[htb]
\caption{The final image ${\hat {\cal Z}}(\theta)$ in the  flattening model  at $\theta=2.32$ and
$\theta=3.07$ for  various values of $p_0$. The default model  is chosen to be $m(\theta)=1.0$. The exact values ${\cal  Z}_{\rm pois}(\theta)$
are also listed.   The results of the Fourier method are  denoted as ${\cal Z}_{\rm four}$.}
\label{table:V&Z&deltaZ} 
\begin{center}
\begin{tabular}{c||r@{.}lr@{.}l r@{.}l}
\hline
\hline
$p_0$ &  \multicolumn{2}{c}{ ${\cal  Z}_{\rm
pois}(2.32)$} &
\multicolumn{2}{c}{ ${\cal Z}_{\rm four}(2.32)$ } &
\multicolumn{2}{c}{ $\hat {{\cal Z}}(2.32)$ } \\
\hline
4.5$\times 10^{-2}$  & 9&801$\times 10^{-3}$ & 9&6(1)$\times 10^{-3}$ &
9&66(1)$\times 10^{-3}$  \\
4.5$\times 10^{-3}$  & 1&094$\times 10^{-3}$ & 9&1(1.5)$\times 10^{-4}$ &
8&99(2)$\times 10^{-4}$ \\
1.0$\times 10^{-3}$   & 3&342$\times 10^{-4}$ & 1&5(1.5)$\times 10^{-4}$ &
1&78(8)$\times 10^{-6}$  \\
\end{tabular}
\vspace{3mm}
\begin{tabular}{c||r@{.}lr@{.}l r@{.}l}
\hline
\hline
$p_0$ & 
\multicolumn{2}{c}{  ${\cal  Z}_{\rm pois}(3.07)$ } &
\multicolumn{2}{c}{ ${\cal Z}_{\rm four}(3.07)$ } &
   \multicolumn{2}{c}{$\hat {{\cal Z}}(3.07)$}\\
\hline
4.5$\times 10^{-2}$  &  9&686$\times 10^{-1}$ & 
   9&5(2)$\times 10^{-3}$ &9&521(2)$\times 10^{-3}$ \\
4.5$\times 10^{-3}$  &
   9&772$\times 10^{-4}$ & 7&5(1.7)$\times 10^{-4}$ &
   7&410(6)$\times 10^{-4}$\\
1.0$\times 10^{-3}$   &  2&174$\times 10^{-4}$ & 
-1&5(18.0)$\times 10^{-5}$ &3&6(2)$\times 10^{-6}$ \\
\end{tabular}
\end{center}
\end{table}
Having distinguished two kinds of flattening, we investigate  whether the MEM can properly reproduce the true one.
We employ the case for $p_0=4.5\times 10^{-2}$, as an example.
Figure~\ref{fig:ZV50c4e-2} displays ${\cal Z}^{(\alpha)}(\theta)$, which is calculated
 by  Eq.~(\ref{eqn:maximumcondition}), for various values of $\alpha$. The default model is chosen to be the constant,
 $m(\theta)=1.0$.   We find that ${\cal Z}^{(\alpha)}(\theta)$ 
are  independent  of the values of $\alpha$.  They show clear flattening  in agreement with
 the result of the Fourier transform and also with that of the Poisson sum formula,  Eq.~(\ref{eqn:poissonsum}).\par
In order to calculate the final image of the partition function,  we need the probability 
${\rm prob}(\alpha|P(Q),I,m)$ as shown in Eq.~(\ref{eqn:averageofZ2Pa}).
For $m(\theta)=1.0$, ${\rm prob}(\alpha|P(Q),I,m)$ is plotted in Fig.~\ref{fig:PV50c4e-2}.
A peak of ${\rm prob}(\alpha|P(Q),I,m)$ is located at $\alpha\approx 12$. 
The integration to obtain the final image $ {\hat {\cal Z}}(\theta)$ thus becomes   
trivially simple because of the fact that ${\cal Z}^{(\alpha)}(\theta)$ is  independent of $\alpha$ in the relevant range
 for the integration of the procedure 2.
Thus $ {\hat {\cal Z}}(\theta)$ agrees with ${\cal Z}^{(\alpha)}(\theta)$ in Fig.~\ref{fig:ZV50c4e-2}, 
 and the MEM yields  true flattening.
 Note that this situation   is unchanged when   the  Gaussian  default model is used, where  $\gamma$ is varied between  0.1  and  3.0.
\par
With   other choice of the value of  $p_0$, we repeat the same procedure. 
For $p_0=4.5\times 10^{-3}$, the MEM yields  true flattening for the final image  $ {\hat {\cal Z}}(\theta)$ as  plotted in  Fig.~\ref{fig:ZV50c4e-3}. The errors are also included but invisible in the 
figure.  The result of the MEM agrees with the 
behavior  of the  exact partition function (Poisson).
For  smaller values of  $p_0$,   however,  flattening is not reproduced correctly.
This is  displayed in Fig.~\ref{fig:ZV50c1e-3} for $p_0=1.0\times 10^{-3}$; although  $ {\hat {\cal Z}}(\theta)$  in the MEM 
flattens  for large values of $\theta$,  its height  is about 4 times  smaller
than the exact one  (Poisson). 
 This  analysis shows that   there is a  critical value of $p_0$ above which the MEM is 
 effective,  and   it is found that  $p_0$ $\approx 2.0\times 10^{-3}$ in the present case.  
 This critical value  is associated with the magnitude of the error in $P(Q)$ or $\delta$. This will be  discussed    later. 
 The result of the Fourier method, ${\cal Z}_{\rm four}(\theta)$,  is also shown as a comparison  in Fig.~\ref{fig:ZV50c1e-3}.  Here,  ${\cal Z}_{\rm four}(\theta)$  receives large errors for  $2.3 \siml \theta \siml 2.5$. For $\theta \simg 2.5$, 
${\cal Z}_{\rm four}(\theta)$ takes negative values  and are  not plotted in the figure.
Table~\ref{table:V&Z&deltaZ} displays the numerical values  of $ {\hat {\cal Z}}(\theta)$ at two typical values of $\theta$, 2.32 as an  intermediate region   and 3.07 as the one near  $\pi$,  for
three values of $p_0$.    The exact values ${\cal  Z}_{\rm pois}(\theta)$ and ${\cal Z}_{\rm four}(\theta)$
are also listed. 
\subsection{$\delta$ dependence}
In the model discussed above, the applicability of the MEM is deeply associated with the
magnitude of the error  in $P(Q)$. 
In order to investigate  this aspect,  we  vary the value of $\delta$, which controls the magnitude
 of the error in the   mock data. 
 Although it would  not  be realistic  to use data with too small  error corresponding to 
 huge amount of statistics,   it would still be   meaningful    to investigate the limitations of the MEM analysis.   This is beneficial   of using mock data. \par
\begin{figure}
        \centerline{\includegraphics[width=9cm, height=7
cm]{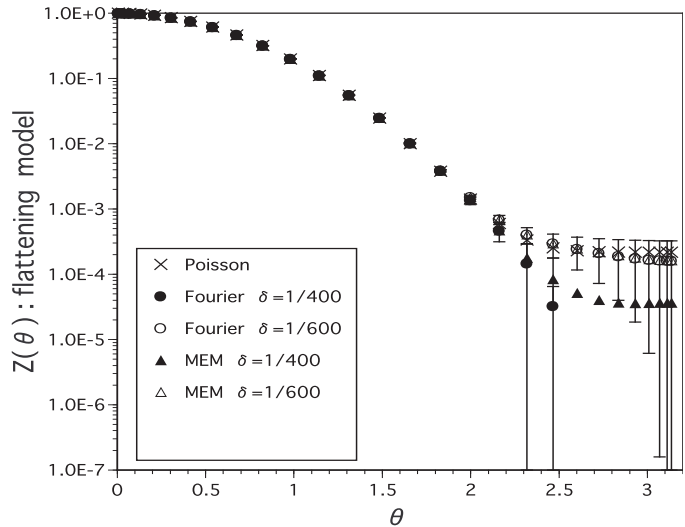}}
\caption{${\cal Z}(\theta)$ in the flattening model  obtained by the numerical Fourier transform  and the MEM
for  $V=50$, $p_0=1.0\times 10^{-3}$. The value of  $\delta$ is chosen to be $1/400$ and $1/600$ for a comparison. Note that ${\cal Z}_{\rm four}(\theta)$ takes negative values for  $\theta\geq  2.5$ in the case 
$\delta=1/400$ and thus these data are not shown.
Error bars  for ${\cal Z}_{\rm four}(\theta)$ with $\delta=1/600$  are   large, while    those of the MEM are invisible.
 }
\label{fig:ZV50c1e-3del}
\end{figure}
 For the flattening model, we   fix the value of  $p_0$ and vary that of  $\delta$. 
 Let us take an example of  $p_0=1.0\times 10^{-3}$, which is   about half   the critical value of
 $p_0$ for the   $\delta=1/400$ case. As already seen in Fig.~\ref{fig:ZV50c1e-3},  both of the methods were not effective   in  reproducing flattening.
We vary  the value of $\delta$ from  1/400 down  to  $1.0\times 10^{-4}$. 
It is found that 
 already for $\delta=1/500$, the MEM reproduces  flattening reasonably well. 
  Figure \ref{fig:ZV50c1e-3del} displays  the results of the MEM analysis and those of the numerical 
Fourier transform  for     $\delta=$  1/400 and 1/600 as a comparison.  The error bars shown in the figure are those of Fourier transform,  while  those for the MEM are invisible.  
For $\delta=1/600$, although the errors  for the Fourier method are large at $\theta\approx \pi$, the central values of ${\cal Z}_{\rm four}(\theta)$ are  approximately equal  to those of the MEM, 
  in contrast to the $\delta=1/400$ case where ${\cal Z}_{\rm four}(\theta)$  takes negative values for $\theta \simg 2.5$.  
  At  $\theta=3.07 $,  the relative error of ${\cal Z}_{\rm four}(\theta)$ goes down to 50 $\%$ for $\delta=1.0\times 10^{-3}$ and 4.6 $\%$ for $\delta=1.0\times 10^{-4}$ while it is about 100$\%$ for $\delta=1/600$.  \par 
  When  $p_0$   is  chosen to be smaller  than $1.0\times 10^{-3}$,  the MEM can recover  correct  images    by taking smaller value of $\delta$ than 1/600. 
  For a given value of $p_0$,     the MEM  becomes effective in  reproducing  correct images, when 
    the  value  of $\delta$ is  reduced  to    the one 
  corresponding to the  precision that $\sum_Q|\delta P(Q)|\approx p_0$.   With  this precision of data,  ${\cal Z}_{\rm four}(\theta)$ still receives much larger  errors than those of the MEM.


 \section{Singular $f(\theta)$ model}
\label{sec:singularf}\par
 \subsection{A mathematical model}
 Although the  flattening   model  develops true flattening,  it     does not reveal a singularity in   $f(\theta)$.
  In order to study  such a  singular behavior mimicking  a  first order phase transition  
  at finite value of $\theta (\neq \pi)$,  we consider  
  a model which utilizes $P(Q)$ obtained by  the inverse Fourier transform of  a  singular $f(\theta)$. 
 Consider a partition function
 \begin{eqnarray}
{\cal Z}(\theta)=
\left\{ \begin{array}{ll}
{\cal Z}_{\rm pois}(\theta) & \mbox {($\theta\leq \theta_{\rm f})$}\\
   {\cal Z}_{\rm f}  & \mbox {$(\theta_{\rm f}<\theta\leq \pi )$, }
  \end{array}
  \right.
\end{eqnarray}
 where ${\cal Z}_{\rm pois}(\theta)$ is the one obtained from the  Poisson sum formula Eq.~(\ref{eqn:poissonsum}) with  $p_0=0$,  and ${\cal Z}_{\rm f}$ is a positive constant ($= {\cal Z}_{\rm pois}(\theta_{\rm f}) $). 
 This partition function is inverse-Fourier-transformed to obtain $P(Q)$:
 \begin{equation}
 P(Q)=\frac 1 \pi \left[ \int_0^{\theta_{\rm f} }d\theta {\cal Z}_{\rm pois}(\theta) \cos \theta Q 
 +\int_{\theta_ {\rm f}} ^\pi d\theta {\cal Z}_{\rm f} \cos \theta Q \right].
  \label{eqn:Pqb}
 \end{equation}
   The singular behavior at $\theta_{\rm f}$ forces  $P(Q)$ to oscillate  for large values of $Q$,   and $P(Q)$ could  take negative values in such a region. 
 Although $P(Q)$ looses the physical  meaning as  a topological charge  distribution for their negative values, we regard such a set of    data of $P(Q)$ as a mathematical   model. What we are concerned with 
 is  rather whether  the singular behavior of ${\cal Z}(\theta)$ is reproduced.\par
    In the numerical calculation, however, the singularity appears  in a approximated way.    
Also,     $P(Q)$ is truncated at an appropriate value of $Q$  satisfying  a  constraint  to keep   the precision of the covariance matrix, which was  described in the subsection~\ref{sec:mock}.  As a consequence, we consider the case where only positive $P(Q)$ are used as well as  
    the one where  negative contribution is also taken into account.
     We use the distributions $P(Q)$, Eq.~(\ref{eqn:Pqb}),    to generate mock data by adding 
   Gaussian noises to them  as  in the flattening  model. We refer to  this as   singular $f(\theta)$ model.
    \par
\begin{figure}
        \centerline{\includegraphics[width=9cm, height=7
cm]{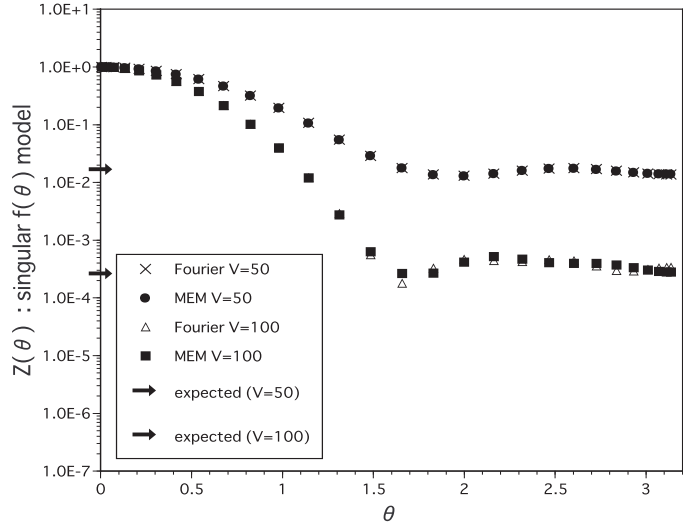}}
\caption{${\cal Z}(\theta)$ obtained by the numerical Fourier transform  and by the MEM
 for $\theta_{\rm f}=\pi/2$ in  the singular $f(\theta)$ model.  The upper (lower) one corresponds to   $V=50 (100)$. The default model is chosen to be the constant 
 $m(\theta)=1.0$. Here, $\delta=1/400$.  Error bars are  displayed but invisible in both of the methods. 
 Arrows indicate the expected heights  of flattening.
 }
\label{fig:ZflV50100m1}
\end{figure}
%
   \begin{figure}
        \centerline{\includegraphics[width=9cm, height=7
cm]{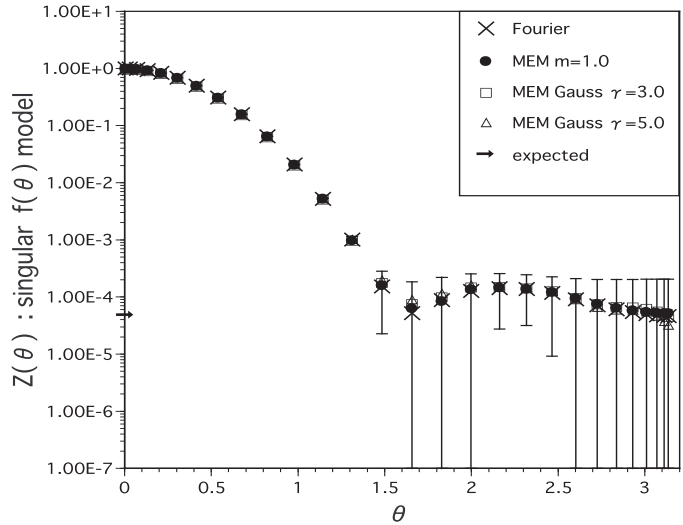}}
\caption{${\cal Z}(\theta)$ obtained by the numerical Fourier transform  and by the MEM
for  $V=120$ and for $\theta_{\rm f}=\pi/2$ in  the singular $f(\theta)$ model.  Here, $\delta=1/400$.  The results with three different default models are shown.   Error bars are of the Fourier method, while  those of the MEM are invisible.  An arrow indicates the expected height of flattening.
 }
\label{fig:ZflV120}
\end{figure}
\begin{figure}
        \centerline{\includegraphics[width=9cm, height=7
cm]{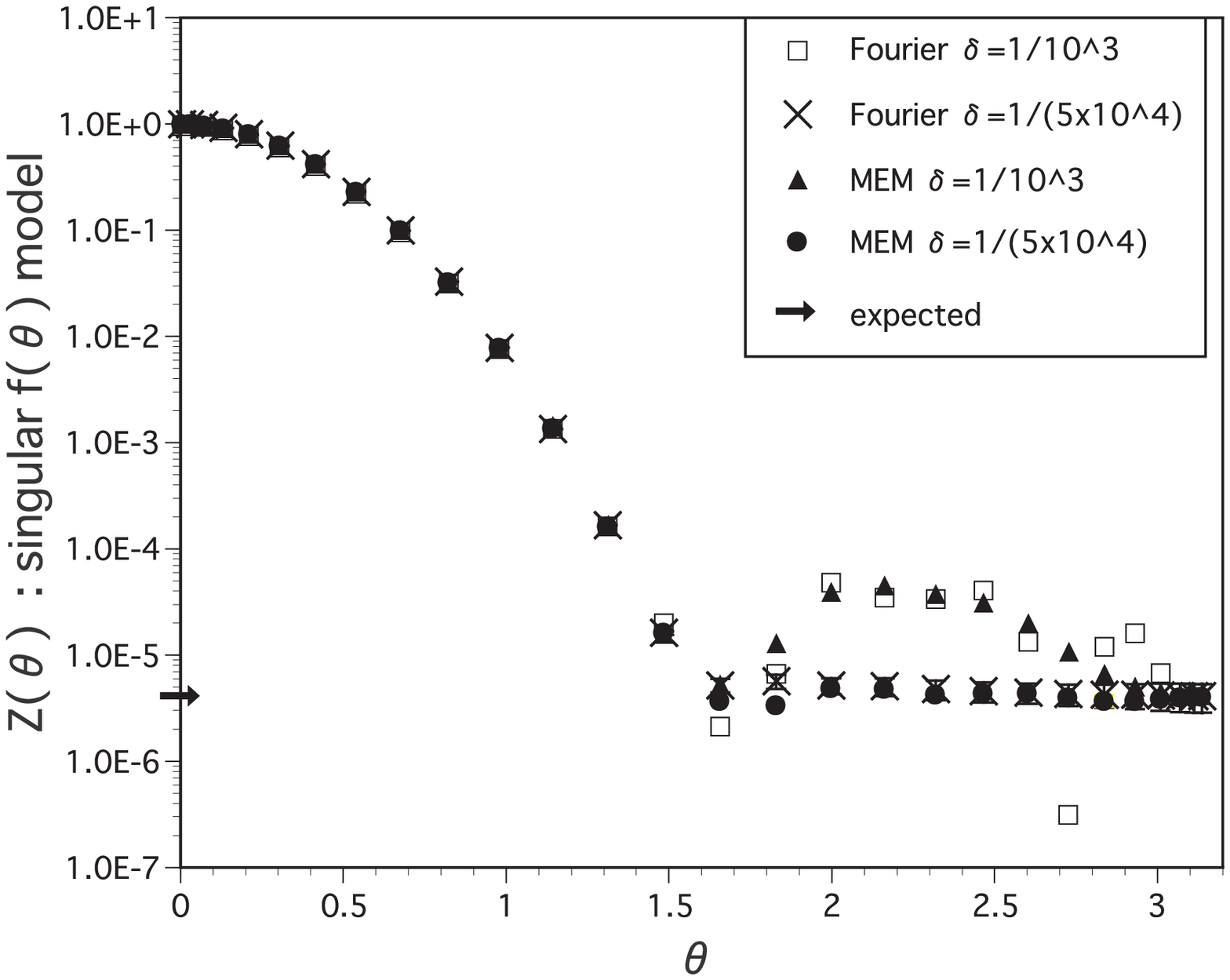}}
\caption{${\cal Z}(\theta)$ in the singular $f(\theta)$ model  obtained by the numerical Fourier transform  and the MEM
for  $V=150$ and $\theta_f = \pi/2$. The results for only  two values of  $\delta$ are  plotted: $\delta=1.0\times 10^{-3}$ and $2.0 \times 10^{-5}$. The constant default model $m(\theta)=1.0$ is used for the MEM analysis. Error bars of  ${\cal Z}_{\rm four}(\theta)$ for  $\delta=1.0\times 10^{-3}$ are not shown, while those for the MEM 
are shown but invisible. 
 }
\label{fig:ZflV150del}
\end{figure}
\subsection{Results}
We are interested in the   behavior  at
 $\theta_{\rm f}$  and  flattening for $\theta\geq \theta_{\rm f}$.
 We fix  $\theta_{f}$ at an appropriate value,  say, $\theta_{\rm f}=\pi/2$ and vary the value of $V$.
 Figure~\ref{fig:ZflV50100m1} displays the result of Fourier transform and  the final image $ {\hat {\cal Z}}(\theta)$ by   the MEM analysis for $V=50$ and 100.
  Here, we truncate $P(Q)$ at some value of  $Q$  so that   $P(Q)$ becomes positive (Q=7 (12) for $V=50     
   ( 100 ) $). 
 For $V=50$, a somewhat acute  bend  at $\theta_{\rm f}$ and flattening for $\theta\simg \theta_{\rm f}$ are  observed. 
Here  we used $m(\theta)=1.0$ as a default model. 
 The height of flattening is in agreement with the input value of  ${\cal Z}_{\rm f}= {\cal Z}_{\rm pois}(\pi/2) $.
 For $V=100$,  singular   behavior at $\theta_{\rm f}$ becomes   clearer, and the height of flattening 
 decreases in agreement with its input value.  As $V$ increases, the bend at $\theta_{\rm f}$ becomes more prominent,  but   the flattening behavior becomes gradually   distorted.  This is because maximal value of $Q$ used for the analysis is restricted   in order to keep the   necessary precision of the covariance matrix as stated  in  subsection \ref{sec:mock}~\cite{rf:ISY}.  \ These features  can also  be seen in  the case  $V=120$ as shown in  Fig.~\ref{fig:ZflV120}.    Here, the height of flattening  agrees with the input value ${\cal Z}_{\rm f}$ indicated by an arrow in the figure.   Figure~\ref{fig:ZflV120} displays  default model dependence of the MEM  results, where  the default models with $m(\theta)=1.0$ and the  Gaussian default with $\gamma=3.0$ and  $\gamma=5.0$ are chosen.  For two values of $\theta$,  the values of  ${\cal Z}(\theta)$  are listed in Table~\ref{table:Bth&Z&deltaZ}.   It  is found  that  the final images of ${\cal Z}(\theta)$  are  independent of  these default models, and 
the magnitudes of the errors are almost the same,  1$-$ 3 $\%$.  
 Figure~\ref{fig:ZflV120} also displays    the result of the Fourier method ${\cal Z}_{\rm four}$, 
  where   quite large errors are generated  for this value of $V$.    In the case of $V=120$, $P(Q)$ is truncated  at $Q=15$. \par
 As $V$ increases further, the height of flattening  decreases, and  the image is  much more affected  by the  error of $P(Q)$.  Here also,   the  error of $P(Q)$ puts   limitations to  the applicability of the MEM as in   the flattening model.  For $\theta_{\rm f}=\pi/2$, it is found that  the MEM cannot be applied already at   $V=150$, where the height of flattening of the image $ {\hat {\cal Z}}(\theta)$ comes out  about one order larger than  the   input value ${\cal Z}_{\rm f}$, although an acute   behavior is still seen  at $\theta_{\rm f}$. 
 This  will be discussed  in the following subsection.
 \begin{table}[htb]
\caption{$\hat {{\cal Z}}(\theta)$ in the singular $f(\theta)$ model  at $\theta=2.32$ and
$\theta=3.07$ for  various default models.  Here, $V=120$. }
\label{table:Bth&Z&deltaZ}
\begin{center}
\begin{tabular}{c||r@{.}lr@{.} lr@{.}lr@{.}l}
\hline
\hline
$\theta$ &  \multicolumn{2}{c}{ $\hat {{\cal Z}}_{m=1}$ } &
\multicolumn{2}{c}{ $\hat {{\cal Z}}_{\gamma =3.0}$ } &
\multicolumn{2}{c}{ $\hat {{\cal Z}}_{\gamma =5.0}$ } &
\multicolumn{2}{c}{  ${\cal Z}_{\rm four}$ } 
\\
\hline
2.32  & 1&39(3)$\times 10^{-4}$ & 1&47(3)$\times 10^{-4}$ &
1&46(3)$\times 10^{-4}$ &   1&4 (1.1)$\times 10^{-4}$ \\
3.07  & 5&31(5)$\times 10^{-5}$ & 5&78(5)$\times 10^{-5}$ &
4&65(5)$\times 10^{-5}$ &
   4&7(15.7)$\times 10^{-5}$ \\
\end{tabular}
\end{center}
\end{table}
\subsection{$\delta$ dependence}
By taking advantage of the mock data, we can take very small values of $\delta$.
When the errors in $P(Q)$ are much reduced, the Fourier method provides the results 
with small errors. 
 Let us take an example of  $V=150$ in the singular $f(\theta)$ model, to which 
both of the methods   were not effective  for $\delta=1/400$ as stated  above.
We choose  $\delta=2.0\times 10^{-5}$ and show the results in Fig.~\ref{fig:ZflV150del}.  
Those  for  $\delta=1.0\times 10^{-3}$ are also plotted as a comparison. 
It is found  for $\delta=1.0\times 10^{-3}$ that 
 both of the methods are still  poor in reproducing flattening, where 
the error bars for the results of the Fourier method are not shown in the figure,  because they are too large.
In the case of $\delta=2.0\times 10^{-5}$, on the other hand, a sharp bend at $\theta_{\rm f}=\pi/2$ and flattening at $\theta \geq \pi/2$ are clearly seen. Both of  the methods provide consistent 
results within the errors.  Here $P(Q)$ is truncated at $Q=19$ and only positive $P(Q)$  are used. \par
So far we have presented the results  with only positive $P(Q)$ in the singular $f(\theta)$ model. 
Let us comment about the negative $P(Q)$ in the  model. 
For $V=150$ in Fig.~\ref{fig:ZflV150del}, for example,  we employed positive $P(Q)$ with $Q=0 - 19$.
 For larger values of $Q$,  $P(Q)$   oscillates and damps 
 with 	a maximal  amplitude of  about $5\times 10^{-8}$.   Negative values are also contained  like 
  $P(Q)=4.814\times 10^{-8},  2.910\times 10^{-8}, -3.010\times 10^{-8}, -2.140\times 10^{-8},  2.696\times 10^{-8}, 1.731 \times 10^{-8}$ and $-2.395\times 10^{-8}$ for $Q=18 - 24$, respectively. By regarding  these values as data and adding Gaussian noise to them,   the resultant  ${\cal Z}(\theta)$   in the Fourier method differs from that in Fig.~\ref{fig:ZflV150del} by at most  2 $\%$ in the region   $ \theta_{\rm f}\siml\theta\leq \pi $,
  while in the MEM   about 6 $\%$ of maximal  difference is found at $\theta=1.8 (> \theta_{\rm f})$.

\section{Summary}
\label{sec:summary}\par
We have considered      lattice field theory with a $\theta$ term,
which suffers from
the complex Boltzmann weight problem or the sign problem in numerical 
simulations.  We have applied the MEM  to the flattening phenomenon of the free energy originating 
from the sign problem.   In our previous analysis, we studied fictitious flattening of $f(\theta)$.
  In the present paper, we have investigated whether the MEM could reproduce true flattening. For this, we have used   mock data  based on the  simple models.
In the flattening model, 
the MEM reproduces true flattening.   This is different from 
what was obtained in (I)~\cite{rf:ISY}  in the fictitious flattening case, where the MEM could yield   monotonically decreasing behaviors  of 
${\cal Z}(\theta)$  in the whole region of $\theta  (0\leq \theta  \leq \pi)$.
Whether the MEM could reproduce true flattening depends on the magnitude of the error.
We have then   investigated how the final images  are  affected by the  errors in  $P(Q)$ by varying the value of $\delta$. 
In the flattening model with $\delta=1/400$, when $p_0$ is as small as $1.0\times 10^{-3}$, both of the results of  the MEM and the 
Fourier method  fail  to reproduce flattening. However, as $\delta$ decreases from $1/400$, an immediate  
 improvement has been  seen   for the MEM.
  For a given value of $p_0$,      the MEM  becomes effective in  reproducing  correct images, when 
    the  value  of $\delta$ is  reduced  to    the one 
  corresponding to the  precision that $\sum_Q|\delta P(Q)|\approx p_0$.   With  this precision of data,  ${\cal Z}_{\rm four}(\theta)$ still receives much larger  errors than those of the MEM.
  \par
 Since the flattening model is free from the singularity, we have  considered a model, 
 the singular $f(\theta)$ model, which mimics a first order phase transition at $\theta (<\pi)$.
 Although this is not a  physical model in the strict sense, it would be useful for studying whether  the MEM  would reproduce  the singular behavior.   Such behaviors  of  the free energy at $\theta_{\rm f}$  are   approximately reproduced depending on the  precision of the data of $P(Q)$.  
 Here also, the MEM is effective when  $\sum_Q|\delta P(Q)|\siml {\cal Z}_{\rm f}$.
 \par 
 The errors  of the final image  in the MEM are  calculated from    the uncertainty of the image. 
   In  the procedure 3   in section~\ref{sec:MEM},   fluctuations  of the image and the correlations among different $\theta$ are properly taken into account according to the probability. However, 
  the smallness of  the uncertainty in the MEM   may not reflect   correct magnitude of the errors.
  Estimate of such  systematic errors  is a task to be done  in the future. 
  On the other hand, 
  the Fourier  method reproduces the results reflecting the errors of $P(Q)$ due  to the propagation of the errors.     This gives  $\sum_{Q} |\delta P(Q)|$ as an estimate of  the magnitude of the errors $|\delta {\cal Z}(\theta)|$ for  any value of $\theta$, which agree approximately  with the results given  in sections~\ref{sec:toymodel} and \ref{sec:singularf}.
 In the limit of  $\delta P(Q)\rightarrow 0$, therefore,  the Fourier  method yields correct behavior,
 whose precision  is   only  affected   by the one   of  computations  employed  in the analysis;  single, double or quadruple precision.  \par
In the present paper, we have concentrated  on the issue whether  the MEM  would be effective with  true flattening by employing mock data.  A MEM analysis   based on 
 real Monte Carlo data will  also be  necessary.  The results   of such  analysis for  
  the ${\rm CP}^{N-1}$  model will  be presented  in a forthcoming  paper. 

\section*{Acknowledgments}
The authors are grateful to  K.  Kanaya  for useful discussion.
This work is supported in part by  Grants-in-Aid for  Scientific Research (C)(2) of 
the  Japan Society for the Promotion of Science  (No. 15540249)
and of the Ministry of Education  Science, Sports and Culture  (No.'s 
13135213 and 13135217).
 Numerical calculations are partly performed on the computer 
at  Computer and Network  Center, Saga University.
%
     
\end{document}